\documentstyle[12pt,epsf]{article}
    \def\hth/#1#2#3#4#5#6#7{{\tt hep-th/#1#2#3#4#5#6#7}}
\def\midinsert{\@midtrue\@ins}

\ifx\epsfbox\UnDeFiNeD
\def\figin#1{\vskip2in}
\else\def\figin#1{#1}
\fi
\def\Fig#1{Fig.~\the\figno\xdef#1{Fig.~\the\figno}\global\advance\figno
 by1}
\makeatletter
\@addtoreset{equation}{section}
\makeatother
\renewcommand{\baselinestretch}{1.1}

\newcommand{\equ}[1]{(\ref{#1})}
\newcommand{\be}{\begin{equation}}
\newcommand{\ee}{\end{equation}}
\newcommand{\eel}[1]{\label{#1}\end{equation}}
\newcommand{\bea}{\begin{eqnarray}}

\newcommand{\eea}{\end{eqnarray}}
\newcommand{\eeal}[1]{\label{#1}\end{eqnarray}}
\newcommand{\beac}{\begin{equation}\begin{array}{rcl}}
\newcommand{\eeacn}[1]{\end{array}\label{#1}\end{equation}}

\newcommand{\CC}{{\cal C}}

\newcommand{\del}{\partial}

\newcommand{\non}{\nonumber}

\newcommand{\journal}[4]{{\em #1~}{\bf #2}\,(19#3)\,#4}

\newcommand{\pr}{\journal {Phys. Rev.}}

\newcommand{\prl}{\journal {Phys. Rev. Lett.}}

\newcommand{\np}{\journal {Nucl. Phys.}}
\newcommand{\pl}{\journal {Phys. Lett.}}
\newcommand{\mpl}{\journal {Mod. Phys. Lett.}}

\begin{document}
%
\begin{titlepage}
\topskip0.5cm
\hfill\hbox{UCSBTH-96-22}\\[-1.4cm]
\flushright{\hfill\hbox{hep-th/9609059}}\\[3.3cm]
\begin{center}{\Large\bf
On the Moduli Space of $N\!=\!2$ Supersymmetric \\$G_2$ Gauge Theory\\
[2cm]}{\large
Karl Landsteiner\footnote{karll@cosmic1.physics.ucsb.edu}, John M.
Pierre\footnote{jpierre@sbphy.physics.ucsb.edu},
and Steven B. Giddings\footnote{giddings@denali.physics.ucsb.edu}
\\[1.2cm]}
\end{center}
\centerline{Department of Physics, University of California, Santa
Barbara, CA 93107, USA}
\vskip2.cm
\begin{abstract}
We apply the method of confining phase superpotentials to $N\!=\!2$
supersymmetric Yang-Mills theory with the exceptional gauge group $G_2$.
Our findings are consistent with the spectral curve of the periodic Toda
lattice, but do not agree with the hyperelliptic curve suggested previously
in the literature.  We also apply the method to theories with fundamental
matter, treating both the example of $SO(5)$ and $G_2$.
\end{abstract}
\vfill
\hbox{September 1996}\hfill\\
\end{titlepage}
%
%
\renewcommand{\baselinestretch}{1.2}\large\normalsize
%
%
Since the seminal work of Seiberg and Witten \cite{sw} we know that the
exact quantum moduli space of $N\!=\!2$ supersymmetric gauge theories
in the Coulomb phase coincides with the moduli space of a particular
abelian variety. In many cases it has been found that this abelian
variety can actually be obtained as the Jacobian of a Riemann surface.
In the simplest and most regular cases it has been shown that these
Riemann surfaces are given by hyperelliptic curves. To be precise,
hyperelliptic curves have been found for the theories with classical
gauge groups $A_n$, $B_n$, $C_n$, and $D_n$ and various matter fields
transforming in the fundamental representation \cite{everybody}.
Hyperelliptic curves have also been suggested for the exceptional gauge
groups $G_2, F_4, E_{6,7,8}$ \cite{ds2,aam,aag}. Clearly the
construction of these hyperelliptic curves was based on an {\it ad hoc}
ansatz which could be justified by performing certain highly
non-trivial consistency checks such as reproducing the correct
weak coupling monodromies or checking the number of special points in the
moduli space where the supersymmetries can be broken down to $N\!=\!1$.

On the other hand, a somewhat more systematic approach has been opened
by the observation that $N\!=\!2$ supersymmetric gauge-theories are
closely related to integrable systems. In particular it has been shown
first by Gorsky et. al. \cite{rus} that the curves for N=2 SYM with
gauge groups $SU(N)$ appear also as the spectral curves of integrable
systems, namely the periodic Toda lattices. This has been generalized
by Martinec and Warner \cite{mw} to the case of pure gauge theories and
arbitrary gauge group. Further Donagi and Witten \cite{dw} gave an
integrable system for the $SU(N)$ theory with one massive
hypermultiplet transforming in the adjoint representation of the gauge
group. Moreover it has been shown recently that these integrable
structures arise naturally in string theories \cite{klmvw}.

In the case of the classical gauge groups it is fairly easy to make a
connection between the spectral curves of the integrable systems and
the hyperelliptic ones. In fact, all the spectral curves for the groups
$A_n, B_n, D_n$ are themselves hyperelliptic and can be brought into
the standard form $y^2 = \prod (x-e_i)$ by a simple change of
variables. In the case of $C_n$ this can be done after modding out a
$Z_2$ symmetry. Thus for the classical gauge groups the spectral curves
of the integrable systems agree with the prior findings of
\cite{everybody}. For exceptional gauge groups on the other hand the
spectral curves are definitely not hyperelliptic. Also it is not known
if the special Prym, which is the physically relevant subspace of the
Jacobian, could be embedded in the Jacobian of a hyperelliptic curve.
Thus no direct correspondence between the spectral curves of \cite{mw}
and the hyperelliptic ones suggested in \cite{ds2,aam,aag} has been set
up until now. The aim of this paper is to tackle this question. 

In particular we will apply the method of confining phase
superpotentials in order to derive information about the exact quantum
moduli space of $N\!=\!2$ supersymmetric Yang-Mills theory with gauge
group $G_2$. We can then actually try to match this information with
what one obtains from the Riemann surfaces.

That the structure of massless monopoles in supersymmetric gauge
theories with a Coulomb phase can be obtained from effective
superpotentials has originally been shown in the case of $SU(N)$ in
\cite{efgir}. Recently this approach has been generalized to the other
classical gauge groups as well \cite{ty}. The starting point in these
constructions is $N\!=\!2$ gauge theory broken down to $N\!=\!1$ by
adding a tree-level superpotential $W_{tree} = \sum_k g_k u_k $, where
$u_k$ are gauge invariant variables built out of the adjoint Higgs
field of the $N\!=\!2$ vector multiplet. In a vacuum in which the gauge
group is enhanced to $SU(2)\times U(1)^{r-1}$, with $r$ the rank of the
group, the low energy regime is governed by an $N\!=\!1$ $SU(2)$ theory
and $r-1$ decoupled photons. One then assumes that the exact low energy
superpotential is simply given by the tree-level superpotential plus
the effective superpotential of the confining $SU(2)$ theory. In this
way one derives the quantum deformations of the vacuum expectation
values of the $u_k$.

Let us start with some introductory remarks about the group
$G_2$.\footnote{$N\!=\!1$ theories with gauge group $G_2$
and matter in the fundamental  
representation have been previously considered in \cite{gp}.} It can be
characterized as the subgroup of $SO(7)$ which leaves one element of
the spinor representation invariant. It is of rank two but a convenient
basis of its Cartan subalgebra can be given by three elements satisfying
one relation. Thus in
$N\!=\!2$ gauge theory the flat directions can be parametrized by three
complex numbers $e_i$, $i\in\{1,2,3\}$, obeying the constraint $\sum
e_i = 0$. The Weyl group is the dihedral group $D_6$. The group action on the
$e_i$ includes
permutations and the simultaneous sign flip of all three elements.
Gauge invariant variables are chosen to be
\bea
u = \frac 1 2 \sum_{i=1}^3 e_i^2,\qquad & & v = \prod_{i=1}^3 e_i^2\,.
\eeal{casimirs}
If $\Phi$ is a matrix in the fundamental representation of $G_2$ and
lies in the Cartan subalgebra we can take $u=\frac 1 4 tr (\Phi^2)$ and $v
= \frac 1 6 tr (\Phi^6) - \frac{1}{96} (tr(\Phi^2))^3$. The moduli
space of the classical gauge theory is described by the 
characteristic polynomial 
\be
P(x) = \frac 1 x det(x-\Phi) = \prod_{i=1}^3 (x^2-e_i^2)\,.
\eel{classicalpoly}
Dividing by $x$ takes into account that the seven dimensional
representation contains a zero weight. An important feature is that the
non-zero weights of the fundamental representations coincide with the
short roots. Thus we can chose a triple of short roots satisfying
$\sum \alpha^s_i = 0$ and the above definition of $u$ and $v$ corresponds to 
$e_i = \vec{\Phi}\cdot\vec{\alpha^s_i}$.
Since the root lattice of $G_2$ is self-dual we could equally well have
chosen to define the $e_i$'s through the long roots; the right
hand side of \equ{classicalpoly} would still be invariant under the Weyl
group. The gauge invariant variables are then redefined according
to $u \rightarrow 3 u$ and $v \rightarrow - 27 v + 4 u^3$. Further if
we look at the classical discriminant of the polynomial
\equ{classicalpoly} we find\footnote{In this paper we drop all
redundant factors in the
expressions for the discriminants.}
\be
\Delta_{cl} = - 4 u^3 v + 27 v^2\,.
\eel{classicaldisc}
The unique redefinition of the gauge invariant variables leaving the
classical discriminant invariant is given by 
\be
v\rightarrow -v + \frac{4}{27} u^3.
\eel{dualxfm} 
This coincides (up to an overall factor of $27^2$)
with the duality transformation shown above.

Let us now take $N\!=\!2$ Yang Mills theory with gauge group $G_2$ and
add a tree-level superpotential of the form 
\be
W_{tree} = g_1 u + g_2 v\,.
\eel{wtreelevel}
Here $u$ and $v$ are built out of the chiral $N\!=\!1$ multiplets
contained in the $N\!=\!2$ vector field. This leads to vacua with
$e_i=0$ and unbroken $G_2$ or to vacua with unbroken $SU(2)\times U(1)$
where two of the $e_i$ coincide (all three being different from
zero)\footnote{Vacua with one of the $e_i$ being zero would also
correspond to unbroken $SU(2)\times U(1)$, however this time 
with the $SU(2)$ factor
embedded into the short roots. Our special choice of superpotential,
which is motivated by \equ{classicalpoly} does not allow for these
vacua to occur.}. More precisely we find $e_1 = e_2 = e = (-
\frac{g_1}{4g_2} )^{\frac 1 4}$ and $\Phi_{cl} =
diag(e, e,-2e,-e,-e,2e,0)$. The low energy theory is governed by a
confined $N\!=\!1$ $SU(2)$ theory whose scale $\Lambda_L$ is related to
the high energy $G_2$ scale scale $\Lambda_H$ by the scale matching
relation \cite{kss}
\be
\Lambda_H^8\, (3 e^2)^{-2}\, (m_{ad})^{2} = \Lambda_L^6\,.
\eel{scalematch}
Here the first factor arises from matching at the scale where the 
W bosons in the coset $G_2/SU(2)$ become massive, 
and the other from matching at the mass scale of the 
$N\!=\!1$ $SU(2)$ adjoint
multiplet. This mass can be computed in the following manner. We
first rewrite the superpotential in terms of $tr(\Phi^2)$ and
$tr(\Phi^6)$ and then vary $\Phi \rightarrow \Phi + \delta \Phi$ where
$\delta \Phi$ lies entirely in the direction of the unbroken $SU(2)$
\bea
\delta^2 W_{tree} &=& \frac{g_1}{4} tr(\delta\Phi^2) + \frac{g_2}{6}
\left[ 15 tr(\Phi^4\delta\Phi^2) - \frac{3}{16} tr(\Phi^2)^2
tr(\delta\Phi^2) - {3\over 4} tr(\Phi^2) tr(\Phi \delta\Phi)^2\right]  \non\\
&=& \frac 1 2 m_{ad} tr(\delta\Phi^2)\,.
\eea
Evaluating this expression at $\Phi_{cl}$ one finds $m_{ad} = \frac 3 2
g_1$. 

The $N\!=\!1$ $SU(2)$ theory confines through gaugino condensation
such that at low energies we are lead to take as superpotential
\be
W_L = W_{tree} \pm 2 \Lambda_L^3 =
 W_{tree} \pm 2 \sqrt{-g_1 g_2} \Lambda_H^4\,.
\eel{Wexact}
The quantum corrections of the gauge invariant variables 
are then given by
\bea
\langle u \rangle 
&=& \frac{\del W_L}{\del g_1} = 3 e^2 \mp \frac{1}{2e^2} \Lambda_H^4\,\non\\ 
\langle v \rangle 
&=& \frac{\del W_L}{\del g_2} = 4 e^6 \pm 2 e^2 \Lambda_H^4\,.
\eea
Upon eliminating $e$ we find 
\be
\Delta_{\pm} = 27 v^2 - 4 u^3 v \mp 36 \Lambda_H^4 u v - 4 \Lambda_H^8
u^2 \mp 32 \Lambda_H^{12} = 0\,,
\eel{disc}
as condition for a vacuum with massless monopole or dyon. 

We observe
that $\Delta_+$ and $\Delta_-$ intersect transversally in four points 
\be
(u,v) = (e^{ \frac{i\pi n}{2} } \,2\,\sqrt[4]{3}\, \Lambda_H^2 , -e^{
\frac{i 3 \pi n}{2} }\, \frac{4 \Lambda_H^6}{9 \sqrt[4]{3}} ),\qquad 
n=0,\ldots, 3\,.
\eel{n=1}
At these points mutually local dyons become massless. Their
multiplicity equals precisely the number of supersymmetric ground
states of $N\!=\!1$ $G_2$ Yang Mills theory.
In addition there are two points in each $\Delta_+$ and $\Delta_-$
where $\Delta_\pm,\del_u \Delta_\pm, \del_v\Delta_\pm$ and the Hessian
of the second derivatives vanish. They are at
\be
(u,v) = (e^{\frac{i\pi n}{2}} \sqrt{6} \Lambda_H^2, - e^{\frac{i 3
\pi n}{2}} \frac{2\sqrt{2}}{3\sqrt{3}} \Lambda_H^6),\qquad n=0,\ldots, 3\,.
\eel{nonlocal}
%
Here one expects mutually non-local dyons to become massless and thus
produce a superconformal field theory \cite{ad,apsw}. As one sees most
easily from fig.~1, each classical singular line is doubled in
the product $\Delta_+\cdot\Delta_-$ corresponding to the appearance of
a massless monopole-dyon pair in the quantum case. These facts lead us
to conjecture that the complete quantum discriminant of $G_2$ is given
by $\Delta_+\cdot\Delta_-$.
\begin{figure}[ht]
\hbox to \hsize{\hss
\epsfysize=10cm
\epsffile{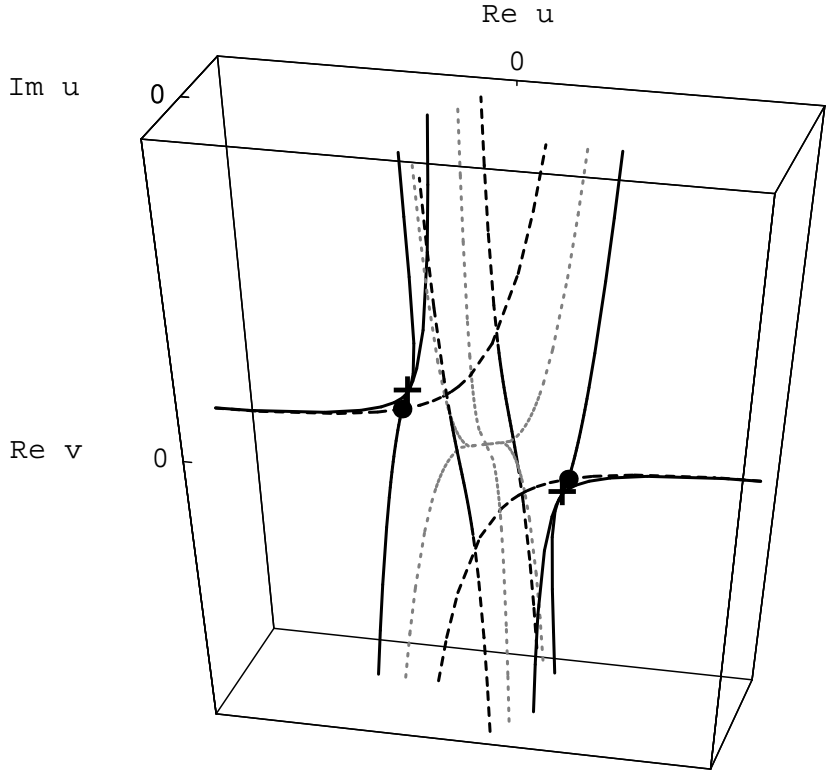}
\hss}

\qquad\qquad\parbox{5in}{\caption{{\tenrm
The $Im\, v=0$ hypersurface of the $G_2$ moduli space.
The light dotted curve is given by the vanishing of the classical
discriminant $\Delta_{cl}=0$.  Classically the $v=0$ plane is also
a singular surface.  The heavy solid and dashed curves are the singular
surfaces $\Delta_{\pm}=0$ where dyons become massless.  Points marked
with a `$\bullet$' indicate transversal intersections where two
mutually local dyons become massless, while at points marked with a
`${\bf +}$' mutually non-local dyons become massless. (Other such points occur
at $Im\, v \neq 0$.)}}\label{fig1}}
\end{figure}

Let us now compare this information with that contained in the
Riemann surfaces of \cite{mw} and \cite{ds2,aam}. We start with the curve
from the integrable system. It is given by the expression
\be
\CC_{int} = 3(z-\frac{\mu}{z})^2 - x^8 + 2 u x^6 - \left[ u^2 + 6
(z+\frac{\mu}{z}) \right] x^4 + \left[v + 2 u (z+\frac{\mu}{z})\right] x^2 =
 0\,.
\eel{integrable}
As explained in \cite{mw} one can view this as an eight-sheeted 
cover of the $z$-plane. 
The parameter $\mu$ is related to the scale $\Lambda_H$. We are
interested in its singularities and therefore look at
\be
\frac{\del \CC}{\del z} =
{2\over z^{3}}(-\mu + z^2)(3\mu + u x^2 z - 3 x^4 z + 3 z^2)=0\,.
\ee
This gives rise to two inequivalent branches\footnote{That we find two
copies of the moduli space is a peculiarity of $G_2$ that may be related to
peculiarities of $G_2$ known in the mathematics literature 
\cite{do}.} of solutions
for $z$. The first one is given by $z= \pm \sqrt{\mu}$. Substituting
this back into \equ{integrable} gives polynomials of sixth order
 in $x$ of the form
\be
P^6_\pm = x^6 - 2 u x^4 + u^2 x^2 \mp 12 \sqrt{\mu} x^2 - v \pm 4 
\sqrt{\mu} u \,.
\ee
The discriminants of these are given by
\begin{eqnarray} \label{polydisc}
\Delta^P_\pm &=& (\mp4\sqrt{\mu}\,u+v) \Delta^{int}_\pm \,,\non\\
\Delta^{int}_\pm &=& \mp 6912 \sqrt{\mu^3} - 144 \mu u^2 \mp 32
\sqrt{\mu}\, u^4 \pm 216 \sqrt{\mu} u v - 4 u^3 v + 27 v^2 \,.
\end{eqnarray}
Both solutions of the second branch generate the same polynomial of eighth 
order
\be
P^8 = 12 x^8 - 12 u x^6 + 4 u^2 x^4 - 3 v x^2 + 36 \mu\,,
\ee
whose discriminant is given by the product of $\Delta^{int}_+$ with
$\Delta^{int}_-$. 
At this point it is important to recall that the construction of the
curves in \cite{mw} was based on the dual (affine) Lie algebra. In
order to compare \equ{polydisc} with \equ{disc} we should therefore
perform a duality transformation as in \equ{dualxfm}.
One finds
then that after setting $\sqrt{\mu} = \frac 1 6 \Lambda_H^4$ the
factors $\Delta^{int}_\pm$ coincide precisely with \equ{disc}! It is
natural then to conjecture that the prefactors in \equ{polydisc} are
`accidental' singularities where ramification points of \equ{integrable}
coincide without physical states becoming massless. 

Furthermore, we can
now classify which superconformal field theory sits at the points
\equ{nonlocal}.  For example, consider $P_-^6$ near the image of the point
$(u_0,v_0) = (\sqrt{6}
\Lambda_H^2, -{2\sqrt2\over 3\sqrt3} \Lambda_H^6)$ where it degenerates.
One can easily show that it takes the form
\be
P^6_- = y^3 - y  \delta u - \delta v\,,
\ee
where $y=x^2 - 2u/3$, $\delta u= \sqrt{8/3} \Lambda_H^2(u-u_0)$, and
$\delta v = v-v_0 + 2 \Lambda_H^4(u-u_0)$ and $u_0, v_0$ are now in the
dual coordinates.  
%
Similar considerations apply to $P^8$. Thus the singularity at this
point is of type $A_2$ and we find a superconformal field theory of the
same type as in \cite{ad}.

Let us turn now to the hyperelliptic curve \cite{ds2,aam}. It can be written
 as
\be
y^2 = \left[x^2 (x^2-u)^2 - v\right]^2 - \Lambda^8 x^4\,,
\eel{hyper}
and its discriminant is given by
\begin{eqnarray} \label{hyperdisc}
v \Delta^h_+ \Delta^h_- &=& v (4 \Lambda^{12} + 8 \Lambda^8 u^2 + 4
\Lambda^4 u^4 - 36 \Lambda^4 u v - 4 u^3 v + 27 v^2)\non\\
& &(-4 \Lambda^{12} + 8 \Lambda^8 u^2 - 4 \Lambda^4 u^4 + 36 \Lambda^4
u v - 4 u^3 v + 27 v^2) \,.
\end{eqnarray}
The first observation is that no redefinition of the gauge invariant
variables can bring this into the form \equ{disc}. That really the
topology of \equ{hyperdisc} is different from what we had before can
also be seen from the fact that $\Delta^h_+$ and $\Delta^h_-$ intersect
each other transversally in eight points which group themselves into
two sets of four ($n=0,\ldots,3$)
\bea
(u,v)_\pm &=& (e^{\frac{i \pi n}{2}} \sqrt[4]{37\pm 14\sqrt{7}} \Lambda^2,
	- e^{\frac{i 3\pi n}{2}} \frac{2}{27} (17\mp 7\sqrt{7})
\sqrt[4]{(37\pm 14\sqrt{7})^3} \Lambda^6) \,.
\eea

Further the discriminant \equ{hyperdisc} has an overall factor of $v$.
Thanks to the curve being hyperelliptic it is not very complicated to
study explicitly the monodromy around this singularity. In order to do
so we go to a scaling limit
where $x=\epsilon z/u$, 
$v=\epsilon^2 \rho$ and $\epsilon \rightarrow 0$.
Then the curve becomes

\be
y^2 \cong \epsilon^4 [(z^2 - \rho)^2 - ({\Lambda^2\over u})^4 z^4].
\eel{yscaled}

\noindent
There are four branch points of this curve, $\pm e_1$ and $\pm e_2$
where

$$ e_1 = \sqrt{\rho \over{1 + ({\Lambda^2 \over u})^2}},
e_2 = \sqrt{\rho \over{1 - ({\Lambda^2 \over u})^2}}.$$

\noindent
Under $v \rightarrow v e^{2 \pi i}$ all the branch points rotate
in the x-plane by $e^{\pi i}$.
Therefore a loop around $v=0$ induces a monodromy which
simply flips the signs of a pair of
dual homology cycles $(\alpha_i, \beta_i)$. We could then choose a basis
such that $(e^D_i,e_i) = \oint_{(\beta_i,\alpha_i)} \lambda$ and find
that say $(e^D_1,e_1) \rightarrow (-e^D_1,-e_1)$ while the others stay
fixed. Already this seems to
be a problem because the Weyl group of $G_2$ does not contain
such an element and moreover it violates the constraint $\sum_i e_i =
0$.  

Let us nonetheless continue to study the physical
implications of this singularity by studying the effective $U(1)$
gauge theory associated with the $(\alpha,\beta)$ cycles.  
To simplify the analysis we
fix $u$ and $\Lambda$ such that $\delta = {\Lambda^4 \over{2 u^2}}$ 
is a small parameter
and use the one form $\lambda = (-4 x^6 + 4 u x^4 - 2 v){dx \over y}$
to solve for the gauge coupling.  We find

$$ a = \int_\alpha \lambda \cong 2 \pi i {\sqrt{v}\over{u}} 
+ {\cal O}(\delta^{2})$$
$$ a_D = \int_\beta \lambda \cong 4 {\sqrt{v}\over{u}} \ln{\delta\over{4}}
+ {\cal O}(\delta)$$  
$$ \tau = {\partial a_D \over{ \partial a}} \cong
{2\over{\pi i}}\ln{\delta\over{4}} + {\cal O}(\delta) $$

\noindent
Note that $\tau$ is scale invariant for fixed 
$\delta$ since it does not depend on 
$\epsilon$,$\rho$, or $v$.  

Now we define $q=e^{\pi i \tau}$ and
substitute into \equ{yscaled} which gives

$$y^2_{eff} = (z^2 - \rho)^2 - 64 q z^4.$$
  
\noindent
Finally, we compare this to the $SU(2)$ curve with $N_f =4$ massless flavors
\cite{sw,as} and
find that they are equivalent in the limit of small $q$.  Indeed
the $SU(2)$ monodromy is just $(a_D, a) \rightarrow (-a_D, -a)$. 
Therefore we conclude that the
physics at the $v=0$ singularity is a non-abelian
Coulomb phase with an enhanced $SU(2)$ gauge symmetry.  
This leads us to an alternative physical interpretation of the 
curve \equ{hyper}.
If we start with the $SU(6)$ curve
with $N_f=4$ massless flavors \cite{ho}
\be
y^2 = (x^6 + \sum_{i=2}^{6} s_i x^{6-i})^2 - \Lambda^8 x^4
\ee
and make a restriction of the moduli space to 
($s_1 = s_3 = s_5 = 0$, $s_2 = -2 u$, $s_4 = u^2$, $s_6=-v$), we
get the same curve as the $G_2$ ansatz.  Now the physics at
$v=0$ has a clear interpretation as a point on the mixed
Higgs-Coulomb branch where classically there is an unbroken
$SU(2)$ gauge symmetry with $4$ massless flavors which is
expected to remain quantum mechanically \cite{aps}.

This analysis together with the fact that the $G_2$ ansatz predicts
more than four $N\!=\!1$ vacua (as well as other strong coupling
monodromies which don't have a clear interpretation) raises serious
doubts about the role of the hyperelliptic curve in describing the strong
coupling physics of $G_2$.  In contrast, although not obviously arising
from a hyperelliptic curve, the discriminant \equ{disc} seems to have the
correct properties to describe $G_2$ and in particular exhibits the
expected splitting of the classical vacua into pairs of singularities.

Certainly our analysis was based on the assumption that the
superpotential \equ{Wexact} is exact. As emphasized in \cite{efgir}, 
{\it a priori} one cannot exclude a correction term $W_\Delta$. In practically
all the examples however it proved to be correct to assume
$W_\Delta=0$, the only exception being the case of $SO(2r+1)$ where the
particular form of $W_\Delta$ just amounted to a redefinition of the
gauge invariant variables similar to the duality transformation we
encountered here. We therefore feel confident that the superpotential
\equ{Wexact} represents the physics correctly.

We now treat the case of the $G_2$ theory with matter
hypermultiplets in the fundamental representation. Before doing so
however, it is instructive to consider the case of $SO(5)$ with one
hypermultiplet. The reason for this is that both models share the
essential physics and we can compare our results for $SO(5)$ with the
known curve of \cite{ha}. The gauge invariant variables are  $u= e_1^2
+ e_2^2$ and $v=-e_1^2 e_2^2$. The tree-level superpotential is
\be
W_{tree} = {\tilde Q}\Phi{Q} + m {\tilde Q}{Q} + g_1 u + g_2 v\,.
\ee
As before, we want to consider classical vacua that break
$SO(5) \rightarrow SU(2) \times U(1)$ which leads to 
\be
\Phi_{cl}  = \left(\matrix{0&ie&0&0&0\cr -ie&0&0&0&0\cr 0&0&0&ie&0\cr
0&0&-ie&0&0\cr 0&0&0&0&0}\right)\ .
\ee
Straightforward computation shows 
$e^2 = {g_1 \over{ g_2}}$. Since the fundamental of $SO(5)$ decomposes
as ${\bf 2}+\bar{\bf 2}+{\bf 1}$,  
at low energies we find an $SU(2)$ theory with two
matter hypermultiplets of masses $m_+ =m+e$ and $m_- =m-e$. The effective
superpotential at low energies is therefore 
\be
W_L = m_+ V^{12} + m_- V^{34} + X(Pf(V)-\Lambda_L^4) + g_1 u +g_2 v.\,,
\ee
Here $V^{ij} = Q^iQ^j$ is the antisymmetric tensor containing the gauge
invariant bilinears of the $SU(2)$ theory and $X$ is a Lagrange
multiplier \cite{se}. We also need the scale-matching condition which
in this case reads $\Lambda^4_L = 4 g_1 g_2\,\Lambda^4_{SO(5)}$.
Eliminating $V^{ij}$ and $X$ by their equations of motion results in
\be
W_L = \frac{g_1^2}{g_2} \pm 2 \sqrt{m^2 - \frac{g_1}{g_2}}\sqrt{g_1
g_2} \Lambda_{SO(5)}^2\,.
\ee
The vev's of the gauge invariant variables are then given by
\bea
\langle u \rangle &=& 2 e^2 \mp
\frac{e\,\Lambda_{SO(5)}^2}{\sqrt{m^2-e^2}} \pm \sqrt{m^2-e^2}\,
\frac{\Lambda_{SO(5)}^2}{e}\,,\\
\langle v \rangle &=& - e^4 \pm \frac{ e^3
\Lambda_{SO(5)}^2}{\sqrt{m^2-e^2}} \pm \sqrt{m^2-e^2}\, e\, \Lambda_{SO(5)}^2
\eea
Upon eliminating $e$ we find that the discriminant is
\bea
\Delta^{N_f=1}_{SO(5)} &=& -16\,{\Lambda_{SO(5)}^{12}}\,{m^4} -
27\,{\Lambda_{SO(5)}^8}\,{m^8} + 36\,{\Lambda_{SO(5)}^8}\,{m^6}\,u -
8\,{\Lambda_{SO(5)}^8}\,{m^4}\,{u^2} +\non\\
& &  {\Lambda_{SO(5)}^4}\,{m^6}\,{u^3} - 
   {\Lambda_{SO(5)}^4}\,{m^4}\,{u^4} +
24\,{\Lambda_{SO(5)}^8}\,{m^4}\,v + 16\,{\Lambda_{SO(5)}^8}\,{m^2}\,u\,v
 +\non\\
& &  36\,{\Lambda_{SO(5)}^4}\,{m^6}\,u\,v -
46\,{\Lambda_{SO(5)}^4}\,{m^4}\,{u^2}\,v + 
   8\,{\Lambda_{SO(5)}^4}\,{m^2}\,{u^3}\,v - 
{m^4}\,{u^4}\,v+ \\
& & {m^2}\,{u^5}\,v + 16\,{\Lambda_{SO(5)}^8}\,{v^2} +
24\,{\Lambda_{SO(5)}^4}\,{m^4}\,{v^2} - 
   64\,{\Lambda_{SO(5)}^4}\,{m^2}\,u\,{v^2} + 
 \non\\
& & 8\,{\Lambda_{SO(5)}^4}\,{u^2}\,{v^2} -
8\,{m^4}\,{u^2}\,{v^2} + 8\,{m^2}\,{u^3}\,{v^2} + 
{u^4}\,{v^2} - 32\,{\Lambda_{SO(5)}^4}\,{v^3} - 
\non\\
& &  16\,{m^4}\,{v^3} + 
16\,{m^2}\,u\,{v^3} + 8\,{u^2}\,{v^3} + 16\,{v^4}\,.\non
\eea
As in the case without matter \cite{ty} this matches the discriminant
of the curve \cite{ha}
\be
y^2 = (x^4 - x^2 u - v)^2 - \Lambda_{SO(5)}^4 x^2(x^2-m^2)\,,
\ee
after rescaling $\Lambda_{SO(5)}$ and 
up to an overall factor\footnote{ The splitting of classical
singularities into monopole-dyon pairs and the number of $N=1$ vacua
expected in the pure YM theory both imply that the $v=0$ singularity 
remains in the full quantum theory in the $SO(5)$ case.  
These are the same principles that lead us
to conjecture that there are no additional factors to
\equ{disc} for $G_2$.} of $v$.  We take
this as convincing evidence for the method. 

It is relatively straightforward to generalize to the case with $N_f$ flavors
using the results of \cite{se}.  We find a low energy superpotential
\be
W_L = {g_1^2 \over g_2} \pm 2 \sqrt{g_1 g_2} 
\sqrt{Pf(\tilde{m})}\, \Lambda_{SO(5)}^{3-N_f},
\ee
where $Pf(\tilde{m}) = \prod_{i=1}^{N_f} (m_i^2 - {g_1 \over g_2})$,
from which we can determine $\langle u \rangle$, $\langle v \rangle$,
and discriminant in the usual way.  For $N_f=2$ we find agreement
with the curve in \cite{ha}.  

Having discussed the case of $SO(5)$ in detail it is straightforward to
treat $G_2$ now. The fundamental of $G_2$ decomposes under $SU(2)$
according to
${\bf 7}\rightarrow {\bf 2}+\bar{\bf 2}+3\cdot {\bf 1}$. 
At low energies we thus have the
same theory as before. The scale matching condition is replaced by
$\Lambda_L^4 = - g_1 g_2 \Lambda_H^6$ and $\Phi_{cl}$ is the same as
without matter. 
For the gauge invariant variables we obtain
\bea
\langle u \rangle &=& 3e^2 \pm \frac{\Lambda_H^3}{4\sqrt{m^2-e^2}} \mp
\sqrt{m^2-e^2} \frac{\Lambda_H^3}{2e^2}\,,\\
\langle v \rangle &=& 4e^6 \pm \frac{ e^4 \Lambda_H^3}{\sqrt{m^2-e^2}}
\pm 2\sqrt{m^2-e^2}\, e^2 \Lambda_H^3\,,
\eea
from which we can in principle
compute the discriminant by eliminating $e$.  For the
case with $m=0$ we find
\bea
\Delta^{N_f=1}_{G2} &=& 8 v (4 u^3 -27 v)^2 + 8748 \Lambda_H^6 v^2 
+ 2160 \Lambda_H^6 u^3 v + 4374 \Lambda_H^{12} v \non\\
& & + 54 \Lambda_H^{12} u^3 + 729 \Lambda_H^{18}
\,.
\eea
It would be extremely interesting to see which kind of Riemann surface
reproduces this result.  Again the generalization to $N_f$ flavors
is given by a low energy superpotential
\be
W_L = i\sqrt{\frac{g_1^3}{g_2}} \pm 2 i \sqrt{g_1 g_2}
\sqrt{Pf(\tilde{m})}\, \Lambda_{G_2}^{4-N_f}
\ee
with $Pf(\tilde{m}) = \prod_{i=1}^{N_f} (m_i^2 - e^2)$ which leads
to
\bea
\langle u \rangle &=& 3e^2 \pm 
\frac{\Lambda_H^{4-N_f}}{4\,\sqrt{Pf({\tilde m})}}
\sum_{i=1}^{N_f}\prod_{j\neq i}(m_j^2 - e^2) 
\mp \sqrt{Pf({\tilde m})} \frac{\Lambda_H^{4-N_f}}{2e^2}\,,\\
\langle v \rangle &=& 4e^6 \pm 
\frac{e^4 \Lambda_H^{4-N_f}}{\sqrt{Pf({\tilde m})}}
\sum_{i=1}^{N_f}\prod_{j\neq i}(m_j^2-e^2)
\pm 2\sqrt{Pf({\tilde m})}\, e^2 \Lambda_H^{4-N_f}\,.
\eea
We can recover results for fewer flavors by taking some quark mass
$m_i \rightarrow \infty$ while keeping 
$m_i (\Lambda_{N_f})^{4-N_f} = (\Lambda_{N_f-1})^{5-N_f}$ held fixed.

\vskip 1cm
{\bf Acknowledgements}

The research of K.L. is
supported in part by the Fonds zur F\"orderung der
wissenschaftlichen Forschung under Grant J01157-PHY and by 
DOE grant DOE-91ER40618.  That of J.M.P. and S.B.G. is partially supported
 by  
DOE grant DOE-91ER40618 and
by NSF PYI grant PHY-9157463.



\begin{thebibliography}{99}
\bibitem{sw} N. Seiberg and E. Witten,
 \np{B426}{94}{19}, \hth/9407087;\\
 N. Seiberg and E. Witten, \np{B431}{94}{484}, \hth/9408099.

\bibitem{everybody} A. Klemm, W. Lerche, S. Theisen, and S. Yankielowicz,
	\pl{B344}{95}{169}, \hth/9411048;\\
P. Argyres and A. Faraggi, \prl{73}{95}{3931}, \hth/9411057;\\
A. Hanany and Y. Oz, \np{B452}{95}{283}, \hth/9505075;\\
P. Argyres, M. Plesser, and A. Shapere, \prl{75}{95}{1699}, \hth/9505100;\\
U. Danielsson and B. Sundborg, \pl{B358}{95}{273}, \hth/9504102;\\
J. A. Minahan and D. Nemeschansky \np{B464}{96}{3}, \hth/9506198;\\ 
A. Brandhuber and K. Landsteiner \pl{358}{95}{73}, \hth/9507008;\\
A. Hanany, \np{B466}{96}{85}, \hth/9509176;\\
P. C. Argyres and A. D. Shapere \np{B461}{96}{437}, \hth/9509175.

\bibitem{ds2} U. Danielsson and B. Sundborg, \pl{B370}{96}{370}, \hth/9511180.
%
\bibitem{aam} M. Alishahiha, F. Ardalan, and F. Mansouri, {\em The
Moduli Space of the Supersymmetric G(2) Yang-Mills Theory}, preprint
IPM-95-117, \hth/9512005.
%
\bibitem{aag} M. R. Abolhasani, M. Alishahiha, and A. M. Ghezelbash,
	{\em The Moduli Space and Monodromies of the $N\!=\!2$
Supersymmetric Yang-Mills Theory with any Lie Gauge Groups}, preprint
IPM-96-144, \hth/9606043.
%
\bibitem{rus} A. Gorsky, I. Kriechever, A. Marshakov, A. Mironov, and A.
Morozov, \pl{B355}{95}{466}, \hth/9505035.
%
\bibitem{mw} E. Martinec and N. P. Warner, \np{B459}{96}{97}, \hth/9509161.
%
\bibitem{dw} R. Donagi and E. Witten, \np{B460}{96}{299}, \hth/9510101.
%
\bibitem{klmvw} A. Klemm, W. Lerche, P. Mayr, C. Vafa and N. Warner,
{\em Self-Dual Strings and $N\!=\!2$ Supersymmetric Field Theory},
preprint CERN-TH-96-95, HUTP-96/A014, USC-96/008, \hth/9604034;\\
K. Landsteiner, E. Lopez and D. Lowe, {\em Evidence for S-Duality in
$N\!=\!4$ Supersymmetric Yang-Mills Theory}, preprint UCSBTH-96-14,
NSF-ITP-96-58, \hth/9606146;\\
C. Gomez, R. Hernandez and E. Lopez, {\em K3 Fibrations and Softly
broken $N\!=\!4$ Supersymmetric Gauge Theories}, preprint
NSF-ITP-96-75, \hth/9608104;\\
S. Nam {\em Integrable Structure in SUSY Gauge Theories and String
Duality}, \hth/9607223;\\
W. Lerche and N. P. Warner, {\em Exceptional SW Geometry from ALE
Fibrations}, \hth/9608183.
%
\bibitem{efgir} S. Elitzur, A. Forge, A. Giveon, K. Intriligator, and E.
Rabinovici, \pl{B379}{96}{121}, \hth/9603051.
%
\bibitem{ty} S. Terashima and S.-K. Yang, {\em Confining Phase  of
$N\!=\!1$ Supersymmetric Gauge Theories and $N\!=\!2$ Massless
Solitons}, preprint UTHEP-340, \hth/9607151.
%
\bibitem{gp} I. Pesando, \mpl{A10}{95}{1871},\hth/9506139;\\
S.B. Giddings and J. Pierre, \pr{D52}{95}{6065}, \hth/9506196;\\
P. Pouliot, \pl{B356}{95}{108}.\hth/9507018.
%
\bibitem{kss} D. Kutasov, A. Schwimmer, and N. Seiberg,
\np{B459}{96}{455}, \hth/9510222.
%
\bibitem{ad} P. Argyres and M. Douglas, \np{B448}{95}{93},
\hth/9505062.
%
\bibitem{apsw} P. C. Argyres, M. R. Plesser, N. Seiberg, and E. Witten,
\np{B461}{96}{71}, \hth/9511154;\\
T. Eguchi, K. Hori, K. Ito, and S.-K. Yang, \np{B471}{96}{430}, \hth/9603002.
%
\bibitem{do} R. Donagi, {\em Decomposition of spectral covers} in {\em
Journ\'ees de Geometrie Alg\'ebrique d'Orsay}, Asterisque vol. 218,
Soc. Math. de France, Paris (1993).
%
\bibitem{ho} A. Hanany and Y. Oz, as in \cite{everybody}.
%
\bibitem{as} P. C. Argyres and A. D. Shapere, as in \cite{everybody}.
%
\bibitem{aps} P. C. Argyres, M. R. Plesser, and N. Seiberg,
\np{B471}{96}{159}, \hth/9603042.
%
\bibitem{ha} A. Hanany, as in \cite{everybody}.
%
\bibitem{se} N. Seiberg, \pr{D49}{94}{6857}, \hth/9402044.
%
\end{thebibliography}
\end{document}